\documentclass[aps,prd,onecolumn,groupedaddress,showpacs,nofootinbib,amssymb,11pt]{revtex4-2}
\usepackage[dvips]{graphicx}
\usepackage{amssymb}
\usepackage{amsmath}
\usepackage{graphicx,,color}
\usepackage{amsfonts}
\usepackage{bm}
\usepackage{cancel}
\usepackage{comment}
\usepackage{floatflt}
\newcommand\be{\begin{equation}}
\newcommand\ee{\end{equation}}
\newcommand\nn{\nonumber \\}
\newcommand\e{\mathrm{e}}

\allowdisplaybreaks[4]

\begin{document}

%\begin{abstrac}
%\end{abstract}

\def\be{\begin{equation}}
\def\ee{\end{equation}}
\def\nn{\nonumber \\}
\def\e{\mathrm{e}}

\title{A reconstruction method for anisotropic universes in unimodular $F(R)$-gravity}

\author{A. Costantini}

\author{E. Elizalde}\email{elizalde@ice.csic.es}

%\author{S.~D.~Odintsov,$^{1,2}$}
%\email{odintsov@ice.cat}

\affiliation{%$^{1)}$ ICREA, Passeig Luis Companys, 23, 08010 Barcelona, Spain\\
Institute of Space Sciences (IEEC-CSIC) Carrer de Can Magrans
s/n, 08193 Cerdanyola del Vallès (Barcelona), Spain\\
}

\begin{abstract}
An extension of unimodular Einsteinian gravity in the context of $F(R)$ gravities is used to construct a class of anisotropic evolution scenarios. In unimodular GR the determinant of the metric is constrained to be a fixed number or a function. However, the metric of a generic anisotropic universe is not compatible with the unimodular constraint, so that a redefinition of the metric, to properly take into account the constraint, need be performed. The unimodular constraint is imposed on $F(R)$ gravity in the Jordan frame by means of a Lagrangian multiplier, to get the equations of motion. The resulting equations can be viewed as a reconstruction method, which allows to determine what function of the Ricci scalar can realize the desired evolution. For the sake of clarity, some characteristic examples are invoked to show how this reconstruction method works explicitly. The de Sitter spacetime here considered, in the context of unimodular $F(R)$ gravity, is suitable to describe both the early- and late-time epochs of the universe history.
\end{abstract}

\maketitle

\section{Introduction}

The Standard Cosmological Model (SCM) provides nowadays a consistent picture of the Universe, at least of its evolution after the inflationary era. It relies on the adoption of General Relativity (GR) as the working theory of gravity, on the Standard Model of (Elementary) Particles (SMP) as the theory that describes the (ordinary) matter content of the universe, and on the cosmological principle, which assumes that (in good accordance with astronomical observations) the Universe is homogeneous and isotropic at cosmic scales. All these assumption are supported by many valuable, independent observations: the spectrum of the cosmic microwave background radiation (CMB), the Big Bang primordial nucleosynthesis (with the specific amounts of primordial elements produced), and the observed large-scale structure (LSS) of the Universe; all together, they provide a quite strong confirmation of the Standard Cosmological Model. However, the observation of the accelerated expansion of the late-time Universe, the necessity to explain the rotation curves of galaxies and some very precise observation of the LSS have led to the necessity of introducing, in addition, two exotic components in the picture, namely dark energy and dark matter. 

Moreover, the need for an inflationary epoch, which has to be necesarilly invoked in order to solve a number of crucial problems of the original Big Bang model, has led to the introduction of a new scalar field (on top of the Higgs boson of the SMP), called the \emph{inflaton}. 

In short, the SCM  outstanding capacity for explaining the more and more accurate astronomical observations does heavily rely on two unknown (dark) components of the energy budget of the universe and on a (somehow misterious) early phase of extremely fast accelerated expansion. Thus, the current cosmological scenario has to deal with basic questions concerning the nature of dark matter and dark energy, and the lack of a natural and universal mechanism for inflation.

One possibility to approach some (or perhaps even all) of these issues is to introduce modifications in the gravitational sector of the theory and to see if one can reliably reproduce the same observed dynamics without the need for any exotic matter/energy component. Modified gravity approaches are indeed quite attractive in order to explain both the dark energy issue and the inflationary paradigm (see, e.g. \cite{Nojiri:2006ri,Nojiri:2010wj,Nojiri:2017ncd, Capozziello:2011et, Cai:2015emx, Capozziello:2009nq, Capozziello:2007ec}), since the early-time and the late-time accelerating expansions can in fact be described, mathematically, in a very similar way, and one can thus expect that the same physical theory underlies both stages \cite{Nojiri:2010wj}. Modified theories of gravity have indeed shown a powerful ability to deal with the fundamental questions of modern theoretical cosmology mentioned above (see, e.g. \cite{Nojiri:2017ncd}).

In the SCM, dark energy appears under the form of the cosmological constant (cc), a fundamental constant of nature introduced by Einstein in 1917, when he used, for the first time, his field equations (obtained two years before) as a tool to describe the universe. As is well known, he used the repulsive effect of the cc in order to obtain a solution depicting a static universe, which was the prevailing conception of the cosmos at that time. This is now paraphrased by some saying that Einstein actually had a visionary insight, also concerning this issue: he introduced a dark energy component for the first time ever (although he did it for an absolutely wrong reason!). The cc, which Einstein himself wrote had ``an still unknown nature'' can now be traded for the contribution of the vacuum energy fluctuations of all fields pervading the universe, in the context of quantum field theory (QFT). At first sight, things seem to perfectly match, at least conceptually. However, the well-known (and awful) problem is that the order of magnitude obtained for the vacuum fluctuations is some 60 to 120 orders of magnitude (depending on the regularization procedure) higher than the observed value of the cosmological constant, which corresponds to the measured acceleration of the universe expansion. 

In GR, same as Einstein did already, the cosmological constant is being added by hand into the Einstein-Hilber action; there is no mechanism to dynamically induce it. From among the plethora of modifications of Einstein's theory, unimodular Einsteinian gravity can be viewed as a useful and quite simple theoretical proposal to explain the presence of such cosmological constant in a geometric way \cite{Alvarez:2005iy,Anderson:1971pn,BUCHMULLER1988292,HENNEAUX1989195,PhysRevD.40.1048,doi:10.1063/1.529283,Finkelstein:2000pg,Alvarez:2006uu,Abbassi:2007bq,Ellis:2010uc,Jain:2012cw,doi:10.1142/S0217732313501307,Kluson:2014esa,Barcelo:2014mua,Barcelo:2014qva,Alvarez:2015sba,Jain:2012gc,Jain:2011jc,Cho:2014taa,Basak:2015swx,Gao:2014nia,Eichhorn:2015bna,Saltas:2014cta}. However, this theory cannot provide a fully satisfactory solution to the problem, as discussed, e.g., in \cite{Padilla:2014yea}. 

In the present paper, we are interested in extending the theory to a covariant formulation of unimodular $F(R)$ gravity, which has richer dynamics, could solve the problem of the cosmological constant, and has the ability to successfully describe both inflationary and bouncing cosmologies \cite{Nojiri:2016ygo}. On top of that, it can also recover the standard Newton law \cite{Nojiri:2016plt} with a similar modification of the usual $F(R)$ gravity, except for an extra term, which can be used to discriminate between unimodular $F(R)$ gravity and the standard $F(R)$ theory, thus rendering the theory falsifiable. We shall consider the same assumption of unimodular Einsteinian gravity, namely the constrained determinant of the metric, and extend the Jordan frame of the $F(R)$ formalism, in order to take properly into account this constraint. While unimodular $F(R)$ gravity has been applied to the Friedmann-Lemaître-Robertson-Walker (FLRW) geometry \cite{Nojiri:2015sfd}, few works exist up to now dealing with anisotropic universes. 

Our aim here is to use this formalism in order to describe the features that the unimodular constraint induces on the anisotropic universes corresponding to $F(R)$ gravity. We will show that the standard metric of a generic anisotropic space-time is not compatible with the unimodular constraint and will be thus compelled to properly fix the metric. Then, we shall derive the corresponding equations of motion, taking always into account the unimodular constraint by means of the use of a Langrange multiplier. We shall finally develop a new reconstruction method (for some related reconstruction methods, see \cite{Nojiri:2006gh,Cognola:2007zu,Nojiri:2009kx,Cognola:2006sp,Elizalde:2008yf}), which has the very interesting feature of being able to realize any given evolution, with a specific Hubble rate, so to explicitly  derive the unimodular $F(R)$ model that yields the given cosmological evolution. 

The paper is organized as follows. In Section II we discuss the general implications of the unimodular constraint on anisotropic universes and develop a reconstruction method for $F(R)$ gravity. In Section III  the reconstruction method is applied to anisotropic metrics with two expansion factors, and it is shown there how to derive the $F(R)$ model that yields the desired cosmological evolution. Finally, our conclusions are presented in  Section IV, the last of the paper.

\section{Anisotropic Universes in Unimodular $F(R)$-Gravity}

In this section we shall provide a generalization of the unimodular $F(R)$ formalism of Ref. \cite{Nojiri:2015sfd} and of its corresponding reconstruction method and adapt it to the description of anisotropic universes. As in standard Einstein-Hilbert unimodular gravity, in unimodular $F(R)$ gravity we will need to fix the determinant of the metric tensor. Throughout the paper, we shall assume the following value for the determinant 
\begin{equation}
    \sqrt{-g} = 1.
    \label{1.1}
\end{equation}
We also assume that the metric describes an homogeneous and anisotropic universe, 
\begin{equation}
    ds^2 = -dt^2 + a(t)^2\sum_{i=1}^3 e^{2\beta_i(t)}\left(dx^i\right)^2.
    \label{1.2}
\end{equation}
Moreover, we impose the following conditions:
\begin{equation}
    \begin{array}{cc}
    \sum_{i=1}^3\Dot{\beta}_i\left(t\right) = 0,  & \sum_{i=1}^3\beta_i\left(t\right) = 0,  \\
    \end{array}
    \label{1.3}
\end{equation}
what can be done without any loss of generality. 

The unimodular constraint \eqref{1.1} is not satisfied by the metric \eqref{1.2}. Thus,  the metric needs to be redefined in a way so that the unimodular metric can be satisfied. We redefine the cosmic time as
\begin{equation}
    d\tau = a(t)^3 dt;
    \label{1.4}
\end{equation}
accordingly, the metric \eqref{1.2} can be rewritten as
\begin{equation}
    ds^2 = -a(t(\tau))^{-6} d\tau^2 + a(t(\tau))^2\sum_{i=1}^3 e^{2\beta_i(t(\tau))}\left(dx^i\right)^2,
    \label{1.5}
\end{equation}
and it can be easily checked that, for the metric \eqref{1.5}, the unimodular constraint \eqref{1.1} is satisfied. 

The unimodular $F(R)$-gravity action in the Jordan frame reads
\begin{equation}
    S = \int d^4x \left\{ \sqrt{-g}\left( F(R) - \lambda \right) + \lambda \right\} + S_{matter},
    \label{1.6}
\end{equation}
where $F(R)$ is a suitable differential function of the Ricci scalar curvature $R$, and $\lambda$ a Lagrange multiplier function; when the action is varied with respect to it, the variation yields the unimodular constraint \eqref{1.1}. Varying the action with respect to the metric, we obtain the following equation of motion: 
\begin{equation}
    G_{\mu \nu}^F = \frac{1}{2}g_{\mu \nu}\left(F(R) - \lambda\right) - R_{\mu \nu}F_R + \nabla_{\mu}\nabla_{\nu}F_R - g_{\mu \nu}\nabla^2F_R,
    \label{1.7}
\end{equation}
where $F_R = \frac{d F(R)}{d R}$.

For the unimodular metric \eqref{1.5}, the connection is 
\begin{equation}
    \begin{array}{ccc}
    \Gamma_{t t}^t = -3K, & \Gamma_{i j}^t = a^8e^{2\beta_i}\left(K + \Dot{\beta}_i \right)\delta_{i j}, & \Gamma_{t j}^i = \Gamma_{j t}^i = \left( K + \Dot{\beta}_i\right)\delta_j^i. \\
    \end{array}
    \label{1.8}
\end{equation}
where the dots indicate derivative with respect to $\tau$, and $K=\frac{1}{a}\frac{d a}{d\tau}$ is the Hubble scale factor for the $\tau$ variable.

We define the Ricci tensor as
\begin{equation}
    R_{\mu \nu} = -\Gamma_{\mu \rho,\nu}^{\rho} + \Gamma_{\mu \nu,\rho}^{\rho} - \Gamma_{\mu \rho}^{\eta}\Gamma_{\nu \eta}^{\rho} + \Gamma_{\mu \nu}^{\eta}\Gamma_{\rho \eta}^{\rho},
    \label{1.9}
\end{equation}
and we find
\begin{equation}
    \begin{split}
        R_{t t} &= -3\Dot{K} - 12K^2 + \sum_i\left( \Dot{\beta}_i^2 \right)^2, \\
        R_{i j} &= a^8e^{2\beta_i}\left[ 6 k \left( K + \Dot{\beta}_i \right) + \Dot{K} + \Ddot{\beta}_i\right]\delta_{i j}, \\
        R &= a^6\left(6\Dot{K} + 30K^2 + \sum_i\left( \Dot{\beta}_i^2 \right)^2 \right).
    \end{split}
    \label{1.10}
\end{equation}
By taking into account Eq. \eqref{1.7}, the $\left(t t\right)$ and  $\left(i j\right)$ components of the equation of motion yield 
%\begin{equation}
%    \begin{split}
%        G_{t t}^F &= -\frac{1}{2}a^{-6}\left( F(R)-\lambda \right) + \left( 3\Dot{K} + 12K^2 + \sum_i\left( \Dot{\beta}_i^2 \right)^2 \right)F_R - 3K \frac{d F_R}{d \tau},\\
%         G_{i j}^F &=a^8e^{2\beta_i} \left\{\frac{1}{2}a^{-6}\left( F(R)-\lambda \right) + \left[ 6 k \left( K + \Dot{\beta}_i \right) + \Dot{K} + \Ddot{\beta}_i\right]F_R + \left(5K+\Dot{\beta}_i\right)\frac{d F_R}{d \tau} + \frac{d^2 F_R}{d \tau^2}\right\},
%    \end{split}
%\end{equation}
\begin{align}
        0 =& -\frac{1}{2}a^{-6}\left( F(R)-\lambda \right) + \left( 3\Dot{K} + 12K^2 + \sum_i\left( \Dot{\beta}_i^2 \right)^2 \right)F_R - 3K \frac{d F_R}{d \tau} +\frac{1}{2}a^{-6}\rho,
    \label{1.11}
\end{align}
\begin{align}
        0=& \frac{1}{2}a^{-6}\left( F(R)-\lambda \right) - \left[ 6 K \left( K + \Dot{\beta}_i \right) + \Dot{K} + \Ddot{\beta}_i\right]F_R + \left(5K - \Dot{\beta}_i\right)\frac{d F_R}{d \tau} + \frac{d^2 F_R}{d \tau^2} + \frac{1}{2}a^{-6}p,
    \label{1.12}
\end{align}
By adding Eqs.\eqref{1.12} with respect to $i$ and using the conditions \eqref{1.3}, one obtains
\begin{equation}
    0 = \frac{1}{2}a^{-6}\left( F(R)-\lambda \right) - \left( 6 K^2 + \Dot{K} \right)F_R +5K\frac{d F_R}{d \tau} + \frac{d^2 F_R}{d \tau^2} + \frac{1}{2}a^{-6}p,
    \label{1.13}
\end{equation}
\begin{equation}
    0 = (6K\Dot{\beta}_i + \Ddot{\beta}_i)F_R + \Dot{\beta}_i\frac{d F_R}{d \tau},
    \label{1.14}
\end{equation}
where Eq. \eqref{1.14} is got by subtracting Eqs. \eqref{1.12} and \eqref{1.13}.

It is possible to integrate Eq. \eqref{1.14} with respect to $\Dot{\beta}_i$, with the result
\begin{equation}
    \Dot{\beta}_i = \frac{C^i a^{-6}}{F_R},
    \label{1.15}
\end{equation}
where the $C^i$'s are constant.

Combining now Eqs. \eqref{1.11} and \eqref{1.13} and eliminating the $\Dot{\beta}_i$'s by using \eqref{1.15}, we get
\begin{equation}
    0=\left( 2\Dot{K} + 6K^2 + \frac{C^2a^{-12}}{F_R^2}\right)F_R + 2K\frac{d F_R}{d \tau} +  \frac{d^2 F_R}{d \tau^2} + \frac{1}{2}a^{-6}\left(\rho + p \right),
    \label{1.16}
\end{equation}
where $C^2 = \sum_i (C^i)^2$.

At this point, if we provide the matter equation of state and the evolution of the scale factor $a = a(t)$, we can derive $\tau$, $K$ and $\Dot{K}$. Eq. \eqref{1.16} becomes then a differential equation for $F_R = F_R\left(\tau\right)$. As the scalar curvature is given by
\begin{align}
    R(\tau) = a(\tau)^6\left( 6\Dot{K} + 30K^2 + \frac{C^2 a^{-12}}{F_R^2} \right),
    \label{1.17}
\end{align}
if we delete $\tau$ by combining $F_R = F_R(\tau)$ and $R = R(\tau)$, then $F_R$ turns into a function of the curvature scalar and can be integrated, to obtain $F(R)$.

\section{Reconstruction Method with Two Scale Factors}

The following metric, with two scale factors 
\begin{align}
    ds^2 = -dt^2 + A(t)^2\left[ (dx^1)^2 + (dx^2)^2 \right] + B(t)(dx^3)^2 \, ,
    \label{1.18}
\end{align}
can be cast in the form of the metric \eqref{1.2}, as 
\begin{align}
    ds^2 &=-dt^2 + a(t)^2 \left\{ \frac{A(t)^2}{a(t)^2}\left[ (dx^1)^2 + (dx^2)^2 \right] +\frac{B(t)^2}{a(t)^2}(dx^3)^2 \right\} = \nonumber \\ 
    &= -dt^2 + a(t)^2 \left\{e^{2\beta_1(t)}\left[ (dx^1)^2 + (dx^2)^2 \right] + e^{2\beta_2(t)}(dx^3)^2 \right\} \, ,
    \label{1.19}
\end{align}
where
\begin{align}
	\beta_1(t) = \log{\left(\frac{A(t)}{a(t)}\right)}, \quad \beta_2(t) = 		\log{\left(\frac{B(t)}{a(t)}\right)}.
	\label{1.20}
\end{align}
To impose the conditions \eqref{1.3}, one needs to define $a(t)$ in a proper way. We start from the second condition of \eqref{1.3}
\begin{align}
	0 = & \beta_1(t) + \beta_2(t) + \beta_3(t)= 2\log{\left(\frac{A(t)}{a(t)}\right)} + \log{\left(\frac{B(t)}{a(t)}\right)} = log{\left( \frac{A(t)^2B(t)}{a(t)^3}  \right)} \nonumber \\ &\Longrightarrow \frac{A(t)^2B(t)}{a(t)^3} =1  
	 \Longrightarrow A(t)^2B(t) = a(t)^3\, ,
	\label{1.21}
\end{align}
while the first one becomes
\begin{align}
    \Dot{\beta}_1(t) = \frac{d}{dt}\log{\left(\frac{A(t)}{a(t)}\right)} = \frac{\Dot{A}}{A} - \frac{\Dot{a}}{a}, \quad
    \Dot{\beta}_2(t) = \frac{d}{dt}\log{\left(\frac{B(t)}{a(t)}\right)} = \frac{\Dot{B}}{A} - \frac{\Dot{a}}{a}
    \label{1.22}
\end{align}
\begin{align}
	0 = \Dot{\beta}_1(t) + \Dot{\beta}_2(t) + \Dot{\beta}_3(t)  = 2\frac{\Dot{A}}{A} + \frac{\Dot{B}}{A} - 3 \frac{\Dot{a}}{a} \Longrightarrow 2\frac{\Dot{A}}{A} + \frac{\Dot{B}}{A} = + 3 \frac{\Dot{a}}{a} \, ,
	\label{1.23}
\end{align}
which is now consistent with both conditions \eqref{1.21}.
In this way, with the scale factor $a(t)$ and the $\beta's$ defined in Eq. \eqref{1.21}.

\subsection{Anisotropic de Sitter Solution}

The metric of an anisotropic de Sitter solution is 
\begin{align}
    ds^2 = -dt^2 + e^{2 H_1 t}\left[ (dx^1)^2 + (dx^2)^2 \right] + e^{2 H_2 t}(dx^3)^2 \, ,
    \label{1.24}
\end{align}
where $H_1$ and $H_2$ are constant.
We can compute the scale factor $a(t)$, as follows
\begin{align}
    a(t) = \left(A(t)^2B(t)\right)^{1/3} = e^{\frac{2H_1+H_2}{3}t} \, ,
    \label{1.25}
\end{align}
while the $\beta's$ are defined as
\begin{align}
    \beta_1 = \log{\left(\frac{A(t)}{a(t)}\right)} = \left(\frac{H_1-H_2}{3}\right) t, \quad \beta_2 = \log{\left(\frac{B(t)}{a(t)}\right)} = \left(\frac{2H_2-2H_1}{3}\right) t.
    \label{1.26}
\end{align}
The cosmic time $\tau$ is redefined as in Eq. \eqref{1.4}, namely 
\begin{align}
    d\tau = a(t)^3 dt \Longrightarrow d\tau = e^{\left(2H_1 + H_2\right)t}dt \Longrightarrow \tau = \frac{1}{2H_1 + H_2}e^{\left(2H_1 + H_2\right)t}.
    \label{1.27}
\end{align}
This relation can be inverted, to obtain $t = t(\tau)$, as
\begin{align}
    t = \frac{1}{2H_1 + H_2}\log{\left[ \left( 2H_1 + H_2 \right)\tau \right]}\, ,
    \label{1.28}
\end{align}
The scale factor $a(t)$ can be written as a function of $\tau$, namely
\begin{align}
    a(t(\tau)) = \left(  3H_0\tau \right)^{\frac{1}{3}}\, ,
    \label{1.29}
\end{align}
and so can the $\beta's$, too, with the result
\begin{align}
    \beta_1 = \frac{H_1 - H_2}{9 H_0}\log{\left( 3H_0\tau \right)}, \quad \beta_2 = \frac{2\left(H_2 - H_1\right)}{9 H_0}\log{\left( 3 H_0\tau \right)},
    \label{1.30}
\end{align}
where $3H_0 = 2H_1+H_2$.

The sum of the squared velocities that appear in \eqref{1.11} can be computed as 
\begin{align}
    \Dot{\beta}_1^2 + \Dot{\beta}_2^2 + \Dot{\beta}_3^2= \frac{2}{27} \left( \frac{H_1 - H_2}{H_0} \right)^2 \frac{1}{\tau^2} \,.
    \label{1.31}
\end{align}
In a quasi-anisotropic approach, we can consider $H_1 \sim H_2$ and the sum of the $\beta$'s squared almost vanish, so that this can properly be considered as a perturbative term. Eq. \eqref{1.16} can be written as
\begin{align}
     \frac{2}{3 \tau} \frac{d F_R}{d \tau} + \frac{d^2 F_R}{d \tau^2} = -\frac{C^2}{81 H_0^4 \tau^4 F_R}\, .
     \label{1.32}
\end{align}
The leading behaviour is obtained when the r.h.s. of \eqref{1.32} vanishes. One finds
\begin{align}
    F_R(\tau) = F_1 + F_2 \tau^{1/3},
    \label{1.33}
\end{align}
where $F_1$ and $F_2$ are constants, to be determined by the initial conditions. When $F_2 = 0$ from the initial conditions, the solution of \eqref{1.32} including the leading correction is given by
\begin{align}
    F_R(\tau) = F_1 - \frac{C^2 \tau^{-2}}{378 H_0^4 F_1}.
    \label{1.34}
\end{align}    
On the other hand, when $F_2 \neq 0$, one finds
\begin{align}
    F_R = F_1 + F_2\tau^{1/3} - \frac{C^2 \tau^{-7/3}}{504 H_0^4 F_2}.
    \label{1.35}
\end{align}
Now, Eq. \eqref{1.17} tells us that
\begin{align}
    R \sim \left\{ \begin{array}{cc}
    12{H_0}^2 + \frac{C^2 \tau^{-2}}{9 H_0^2 F_1^2} & \mathrm{when}\ F_2=0 \\
    12{H_0}^2 + \frac{C^2 \tau^{- 8/3}}{9 H_0^2 F_2^2} & \mathrm{when}\
    F_2\neq 0
    \label{1.36}
    \end{array}
    \right. \, ,
    \end{align}
and, consequently, we get 
\begin{align}
    F_R \sim \left\{ \begin{array}{cc}
    \frac{F_1}{42}\left(54 - \frac{R}{{H_0}^2} \right)  & \mbox{when}\ F_2=0 \\
    F_1 + \frac{F_2 \left( 68 H_0^2 - R \right)}{56 3^{1/3} H_0^2 \left( \frac{F_2^2 H_0^2 \left( R - 12 H_0^2 \right)}{C^2} \right)^{1/8}} & \mbox{when}\ F_2\neq 0
    \end{array}
    \right. \, .
    \label{1.37}
\end{align}
By integrating $F_R$ with respect to $R$, one easily finds the form of $F$, to be
\begin{align}
    F(R) \sim \left\{ \begin{array}{cc}
    F_0 + \frac{1}{42} F_1 \left( 54 R - \frac{R^2}{2 H_0^2} \right)   & \mbox{when}\ F_2=0 \\
    F_0 + F_1 R - \frac{C^2 \left( R-132 H_0^2 \right) \left( \frac{F_2^2 H_0^2 \left(R-12 H_0^2\right)}{C^2} \right)^{7/8} }{105 \sqrt[4]{3} F_2 H_0^4} & \mbox{when}\ F_2\neq 0
    \end{array}
    \right. \, .
    \label{1.38}
\end{align}

Finally, using now Eq. \eqref{1.11}, the unimodular Lagrange multiplier function reads 
\begin{align}
    \lambda(\tau) \sim \left\{ \begin{array}{cc}
    -\frac{C^4}{6804 F_1^3 H_0^6 \tau^4}+\frac{2 C^2}{9 F_1 H_0^2 \tau^2}+\frac{138 F_1 H_0^2}{7}   & \mbox{when}\ F_2=0 \\
    -\frac{C^4 \left(\frac{1}{\tau^{8/3}}\right)^{15/8}}{8505 F_2^3 H_0^6}+\frac{C^2 \left(7 F_1+8 F_2 \left(\frac{1}{\sqrt[8]{\frac{1}{\tau^{8/3}}}}+\sqrt[3]{\tau}\right)\right)}{63 F_2^2 H_0^2 \tau^{8/3}}+18 F_1 H_0^2 & \mbox{when}\ F_2\neq 0
    \end{array}
    \right. \, .
    \label{1.39}
\end{align}

\subsection{Anisotropic Big-Rip Singularity}

Assuming now the following metric
\begin{align}
    ds^2 = -dt^2 + \left( t - t_s \right)^{a_0}\left[(dx^1)^2 + (dx^2)^2\right] + \left( t - t_s \right)^{b_0} \left(dx^3\right)^2\,,
    \label{1.40}
\end{align}
where $a_0, b_0$ are constant, we compute the corresponding scale factor $a(t)$, as 
\begin{align}
    a(t) = \left[ A(t)^2B(t)\right]^{1/3} = \left( t - t_s\right)^{\frac{2 a_0 + b_0}{3}} \,.
    \label{1.41}
\end{align}
Then, we can define the cosmic time $\tau$ as in Eq. \eqref{1.4}, namely
\begin{align}
    d\tau = a(t)^3 dt = \left( t - t_s\right)^{2 a_0 + b_0} dt \Longrightarrow \tau = \frac{\left( t - t_s\right)^{2 a_0 + b_0 + 1}}{2 a_0 + b_0 + 1}\,,
    \label{1.42}
\end{align}
and we can invert this relationship and express $t = t(\tau)$, as follows
\begin{align}
    t - t_s = \left( f_0 \tau \right)^{1/f_0}\,
    \label{1.43}
\end{align}
where $f_0 = 2 a_0 + b_0 + 1$.

We can now compute the $\beta's$, as in Eq. \eqref{1.20}
\begin{align}
    \beta_1(t) = \frac{a_0 - b_0}{3}\log{\left(t-t_s\right)}, \qquad \beta_2(t) = \frac{2b_0 - 2a_0}{3}\log{\left(t - t_s \right) }\, ,
    \label{1.44}
\end{align}
which can be expressed as a function of $\tau$,
\begin{align}
    \beta_1(\tau) = \frac{a_0 - b_0}{3 f_0}\log{\left(f_0 \tau\right)}, \qquad \beta_2(\tau) = \frac{2b_0 - 2a_0}{3 f_0}\log{\left(f_0 \tau \right) }\, .
    \label{1.45}
\end{align}

The sum of the $\Dot{\beta}$'s squared yields
\begin{align}
    \sum_{i=1}^3 = \frac{6\left( a_0 - b_0 \right)^2}{9 f_0^4 \tau^2}
    \label{1.46}
\end{align}
so that Eq. \eqref{1.16} in vacuum can be written as
\begin{align}
    \left( -\frac{2 }{f_0\tau^2} + \frac{6 }{f_0^2 \tau^2} + \frac{6\left( a_0 - b_0 \right)^2}{9 f_0^4 \tau^2} \right)F_R\left( \tau \right) + \frac{2 f_0}{\tau} \frac{d F_R \left(\tau \right)}{d \tau} + \frac{d^2 F_R \left( \tau \right)}{d\tau^2} = 0 \,
    \label{1.47}
\end{align}
which can be solved, the solution being
\begin{align}
    F_R\left( \tau \right) = C_+ \tau^{\alpha_+} + C_- \tau^{\alpha_-}, \qquad \alpha_{\pm} = \frac{1}{2} - \frac{1}{f_0} \pm \frac{\sqrt{+24(a_0 - b_0)^2 + 9f_0^2(f_0^2 + 4f_0 -1)}}{6f_0^2}
    \label{1.48}
\end{align}

The Ricci scalar reads
\begin{align}
    R\left( \tau \right) = \frac{2 \left[ \left( a_0 -b_0 \right)^2 - 9f_0^2 \left( f_0 - 5 \right) \right]}{3 f_0^2} \left( f_0 \tau \right)^{\frac{6-2f_0}{f_0}}\, ,
    \label{1.49}
\end{align}
which can be inverted, to find
\begin{align}
    \tau \left( R \right) = \frac{1}{f_0} \left[ \frac{3/2 f_0^2 R}{\left( a_0 -b_0 \right)^2 - 9f_0^2 \left( f_0 - 5 \right)} \right]^{\frac{f_0}{6-2f_0}} = \frac{1}{f_0}\left( \frac{R}{K} \right)^{\frac{f_0}{6-2f_0}}\,,
    \label{1.50}
\end{align}
and we can rewrite the solutions \eqref{1.48} as
\begin{align}
    F_R = \Tilde{C}_+ \left( \frac{R}{K} \right)^{\frac{ \alpha_+ f_0 }{6-2f_0}} + \Tilde{C}_- \left( \frac{R}{K} \right)^{\frac{ \alpha_- f_0 }{6-2f_0}} \, .
    \label{1.51}
\end{align}
As a  consequence, we get
\begin{align}
    F\left(R\right) = \frac{\Tilde{C}+}{K} \frac{6-2f_0}{6 + f_0(\alpha_+ -2 )} \left( \frac{R}{K} \right)^{\frac{ 6 + f_0(\alpha_+ -2 ) }{6-2f_0}} + \frac{\Tilde{C}_-}{K} \frac{6-2f_0}{6 + f_0(\alpha_+ -2 )} \left( \frac{R}{K} \right)^{\frac{ 6 + f_0(\alpha_- -2 ) }{6-2f_0}}\,,
    \label{1.52}
\end{align}

\subsection{Power-Law Expansion Factor}

Consider now the following metric
\begin{align}
    ds^2 = -dt^2 + \left(\frac{t}{t_0}\right)^{2a_0}\left[ (dx^1)^2 +(dx^2)^2\right] + \left( \frac{t}{t_0} \right)^{2b_0} (dx^3)^2 \,,
    \label{1.53}
\end{align}
where $a_0, b_0$ and $t_0$ are constant. The scale factor can be computed as in Eq. \eqref{1.21}
\begin{align}
    a(t) = \left[ A(t)^2 B(t)\right]^{1/3} =  \left(\frac{t}{t_0}\right)^{\frac{2a_0 + b_0}{3}} = \left(\frac{t}{t_0}\right)^{f_0}\, , 
    \label{1.54}
\end{align}
and the $\beta$'s as in Eq. \eqref{1.32}, to get
\begin{align}
    \beta_1 = \beta_2 = \left( \frac{a_0 - b_0}{3}\right)\log{\frac{t}{t_0}} \qquad \beta_3 =  \left( \frac{2b_0 - 2a_0}{3}\right)\log{\frac{t}{t_0}} \, ,
    \label{1.55}
\end{align}

The cosmic time $\tau$ can be defined as in Eq. \eqref{1.4}, namely
\begin{align}
    d\tau = a(t)^3 dt = \left(\frac{t}{t_0}\right)^{3f_0} \Longrightarrow \tau = \frac{t_0}{3f_0 + 1}\left(\frac{t}{t_0}\right)^{3f_0 + 1},
    \label{1.56}
\end{align}
which can be inverted, to get
\begin{align}
    \frac{t}{t_0} = \left( \frac{3 f_0 + 1}{t_0} \tau \right)^{\frac{1}{3 f_0 + 1}} \, ,
    \label{1.57}
\end{align}
\begin{align}
    a(\tau) = \left(\frac{\tau}{\tau_0}\right)^{h_0}, \qquad \beta_1(\tau) = \beta_2(\tau) = \left( \frac{a_0 - b_0}{9f_0 + 3}\right) \log{\left(\frac{\tau}{\tau_0}\right)}, \qquad \beta_3(\tau) = \left( \frac{2b_0 - 2 a_0}{9f_0 + 3}\right) \log{\left(\frac{\tau}{\tau_0}\right)}\,,
    \label{1.58}
\end{align}
where $\tau_0 = \frac{t_0}{3 f_0 + 1}$ and $h_0 = \frac{f_0}{3 f_0 + 1}$.

The sum of the derivative of the $\beta$'s with respect of the new cosmic time $\tau$ vanishes, as a consequence of the conditions \eqref{1.3}, but the sum of the velocities squared yields
\begin{align}
    \sum_{i = 1}^3 \Dot{\beta}_i^2 = \frac{\left(a_0 - b_0 \right)^2}{9\left(3f_0 + 1 \right)^2}\tau^{-2} \,
    \label{1.59}
\end{align}
and the Hubble factor can also be expressed as a function of $\tau$, as
\begin{align}
    K = \frac{d a(\tau)}{d \tau}\frac{1}{a(\tau)} = \frac{h_0}{\tau}\, .
    \label{1.60}
\end{align}
We can now compute Eq. \eqref{1.16} in vacuum, with the result
\begin{align}
    0 = & \left( \frac{-2 h_0}{\tau^2} + \frac{6 h_0^2}{\tau^2} + \frac{\left(a_0 - b_0 \right)^2}{9\left(3f_0 + 1 \right)^2} \frac{1}{\tau^2}\right) F_R + \frac{2 h_0}{\tau}\frac{dF_R}{d\tau} + \frac{d^2 F_R}{d\tau^2} = \nonumber \\
    & \left(\frac{2}{3} \frac{ a_0^2 - 2 a_0b_0 - 2 a_0 +  b_0^2 -  b_0}{(2a_0 + b_0 + 1)^2} \right)\frac{F_R}{\tau^2} + \frac{2}{3\tau}\left(\frac{2a_0 + b_0}{2a_0 + b_0 + 1}\right)\frac{dF_R}{d\tau} + \frac{d^2 F_R}{d\tau^2}\, .
    \label{1.61}
\end{align}
Solving these equations, one gets the following function
\begin{align}
    F_R(\tau) = C_+\tau^{\alpha_+} + C_-\tau^{\alpha_-}, \qquad \alpha_{\pm} = \frac{2 a_0 + b_0 + 3 \pm \sqrt{9 - 36 a_0^2 + 50 b_0 + 21 b_0^2 + 40 a_0 + 24 a_0b_0}}{6(2a_0 + b_0 + 1)} \,.
    \label{1.62}
\end{align}

The Ricci scalar is, in general, a function of $\tau$ and can be evaluated, for a given metric, with the help of Eq. \eqref{1.17}, as
\begin{align}
    R(\tau) = \left( \frac{\tau}{\tau_0} \right)^{6 h_0}\left( \frac{6 a_0^2 - 4 a_0 + 2 b_0^2 - 2 b_0 + 4 a_0 b_0}{(2 a_0 + b_0 + 1)^2 \tau^2}\right) \, ,
    \label{1.63}
\end{align}
which can be inverted, too
\begin{align}
    \tau(R) = \left( \frac{\tau_0^{h_0}}{C_0} R \right)^{\frac{1}{6h_0 - 2}}, \qquad C_0 = \frac{6 a_0^2 - 4 a_0 + 2 b_0^2 - 2 b_0 + 4 a_0 b_0}{(2 a_0 + b_0 + 1)^2} \, .
    \label{1.64}
\end{align}
The solution \eqref{1.62} can be expressed as a function of $R$, as
\begin{align}
    F_R(R) = C_+  \left( \frac{\tau_0^{h_0}}{C_0} R \right)^{\frac{\alpha_+}{6h_0 - 2}} + C_- \left( \frac{\tau_0^{h_0}}{C_0} R \right)^{\frac{\alpha_-}{6h_0 - 2}} \, ,
    \label{1.65}
\end{align}
and integrated with respect to the curvature scalar, to yield
\begin{align}
    F(R) = C_+\frac{6h_0 -2}{\alpha_+ +6h_0 -2} \left( \frac{\tau_0^{h_0}}{C_0} R \right)^{\frac{\alpha_+}{6h_0 - 2} + 1} + C_-\frac{6h_0 -2}{\alpha_- +6h_0 -2} \left( \frac{\tau_0^{h_0}}{C_0} R \right)^{\frac{\alpha_-}{6h_0 - 2} + 1}\, .
    \label{1.66}
\end{align}

To finish, the Lagrange multiplier can be obtained by using Eq. \eqref{1.13}
\begin{align}
    \lambda(\tau) =& 2\left( \frac{\tau}{\tau_0} \right)^{6h_0} \left[ C_+\left( h_0 - 6 h_0^2 + 5h_0\alpha_+ + \alpha_+^2 + \alpha_+ \right)\tau^{\alpha_+ -2} + C_-\left( h_0 - 6 h_0^2 + 5h_0\alpha_- + \alpha_-^2 + \alpha_- \right)\tau^{\alpha_- -2}\right] + \nonumber\\
    & + \left[ C_+ \left( \frac{6h_0 - 2}{\alpha_+ + 6 h_0 - 2}\right)\tau^{\alpha_+ + 6 h_0 -2 }  + C_- \left( \frac{6h_0 - 2}{\alpha_- + 6 h_0 - 2}\right)\tau^{\alpha_- + 6 h_0 -2 }  \right]\, .
    \label{1.67}
\end{align}

\section{Comparison of the corrections to Newton's law in standard vs unimodular $F(R)$ gravities and an effective equation of state for the last one}

We shall here stress the potential value of using unimodular $F(R)$-gravity for defining a new equation of state, first comparing the corrections it induces to Newton's law with  the corresponding ones in standard $F(R)$-gravity. 

\subsection{Correction to Newton's law in standard $F(R)$-gravity}

Standard $F(R)$-gravity exhibits different corrections to the Newton law with respect to the unimodular one. We can estimate the impact of these corrections in the framework of scalar-tensor $F(R)$-gravity. The action of a general $F(R)$ theory in the Einstein frame is
\begin{align}
     S_E = \frac{1}{k^2}\int d^4x\sqrt{-g}\left( R - \frac{3}{2}g^{\rho \sigma}\partial_{\rho}\phi\partial_{\sigma}\phi - V(\phi)\right),
    \label{1.68}
\end{align}
with $A = R$, $\phi = -\ln{\left(1+f'(A)\right)}$ and $V(\phi) = \frac{A}{F'(A)} - \frac{F(A)}{F'(A)^2}$.

After the scale transformation $g_{\mu \nu} \Longrightarrow e^{\phi}g_{\mu \nu}$,  a coupling of the scalar field $\phi$ with matter appears. Its strength is of the same order of the gravitational coupling $k$; unless the mass of $\phi$ is large, there  large corrections to Newton's law appear. The mass of the scalar field is given by
\begin{align}
    m_{\phi}^2 = \frac{1}{2}\left\{\frac{A}{F'(A)} - \frac{4 F(A)}{F'(A)^2} + \frac{1}{F''(A)} \right\}
    \label{1.69}
\end{align}

By taking in account the model \eqref{1.66}, which can be rewritten in a simpler form, as
\begin{align}
    F(R) = A_+R^{\rho_+} + A_-R^{\rho_-},
    \label{1.70}
\end{align}
we can express the mass  associated to the scalar field, as
\begin{align}
    m_{\phi}^2 = \frac{1}{2}R^2
    \left\{ \frac{A_+(\rho_+ -4)R^{\rho_+} + A_-(\rho_- -4)R^{\rho_-}}{\left( A_+\rho_+R^{\rho_+} + A_-\rho_-R^{\rho_-} \right)^2} + \frac{1}{A_+\rho_+(\rho_+ -1)R^{\rho_+} + A_-\rho_-(\rho_- -1)R^{\rho_-}}    \right\}.
    \label{1.71}
\end{align}
For the sake of simplicity, we take $\rho_+ = \rho_- = \rho$ and $A_+,A_- >0$; thus we can rewrite Eq.\eqref{1.71} as
\begin{align}
    m_{\phi}^2 = \frac{1}{2(A_+ + A_-)}R^{2-\rho}\left[ \frac{\rho - 4}{\rho^2} + \frac{1}{\rho (\rho -1 )} \right].
    \label{1.72}
\end{align}

For very small curvature, as in the Solar System case, the mass of the scalar field $\phi$ becomes small, too, if $\rho$ is in the range $(1,2)$. In the Solar System, $A = R \sim 10^{-61}eV^2$ and if $\rho = \frac{3}{2}$, we obtain
\begin{align}
    m_{\phi}^2 \sim 10^{-\frac{61}{2}}.
    \label{1.73}
\end{align}
Thus, the corrections to Newton's law are too big and the model does not reproduce the correct dynamics in the Solar System, in such situation. 

\subsection{Corrections to Newton's law in unimodular $F(R)$-gravity}

On the contrary, we can prove that we can avoid this situation in the case of unimodular $F(R)$ gravity. For models in the framework of the unimodular $F(R)$ theory we are actually able to reproduce the observed dynamics of the Solar System, in the limit of low curvature.

Eq. \eqref{1.68}, in the case of unimodular $f(R)$-gravity, can be changed as follows
\begin{align}
     S_E = \frac{1}{k^2}\int d^4x\left\{ \sqrt{-g}\left[\left( R - \frac{3}{2}g^{\rho \sigma}\partial_{\rho}\phi\partial_{\sigma}\phi - V(\phi)\right) -\lambda e^{2\phi}\right] + \lambda \right\} + S_{matter}\left(e^{\phi}g_{\mu \nu}, \Psi \right)
    \label{1.74}
\end{align}
The unimodular constraint is now given by $e^{\phi}\sqrt{-g} = 1$. And eliminating the scalar field, the action \eqref{1.74} can be rewritten as 
\begin{align}
     S_E = \frac{1}{k^2}\int d^4x\left\{ \sqrt{-g}\left[\left( R - \frac{3}{32g^2}g^{\rho \sigma}\partial_{\rho}\phi\partial_{\sigma}\phi - V(\frac{1}{4}\ln{(-g)})\right) -\lambda e^{2\phi}\right] + \lambda \right\} + S_{matter}\left((-g)^{1/4}g_{\mu \nu}, \Psi \right).
    \label{1.75}
\end{align}

We now consider perturbations of the metric $g_{\mu \nu}$ around the background metric, assumed to be flat,  as follows, $g_{\mu \nu} = \eta_{\mu \nu} + h_{\mu \nu}$. We thus find that
\begin{align}
    \sqrt{-g}R \sim -\frac{1}{2}\partial_{\lambda}h_{\mu \nu}\partial^{\lambda}h^{\mu \nu} + \partial_{\lambda}h^{\lambda}_{\mu}\partial_{\nu}h^{\mu \nu} - \partial_{\mu}h^{\mu \nu}\partial_{\nu}h + \frac{1}{2}\partial_{\lambda}h\partial^{\lambda}h,
    \label{1.76}
\end{align}
where $h$ is the trace of the tensor field $h_{\mu \nu}$. Because of the flat background choice, we find $V(0) = V'(0) = 0$, and we may write down the potential $V \sim \frac{1}{2}m^2h^2$. Thus, we can write down the linearized action \eqref{1.75}, and varying it with respect to $h_{\mu \nu}$, we obtain the following equations:
\begin{align}
    \partial_{\lambda}\partial^{\lambda}h_{\mu \nu} - \partial_{\mu}\partial^{\lambda}h_{\lambda \nu} - \partial_{\nu}\partial^{\lambda}h_{\lambda \mu} + \partial_{\mu}\partial_{\nu}h + \eta_{\mu \nu}\partial^{\rho}\partial^{\sigma}h_{\rho \sigma} - \frac{13}{16}\nu_{\mu \nu}\partial_{\lambda}\partial^{\lambda}h - m^2\eta_{\mu \nu}h = T_{\mu \nu} - \frac{1}{4}\eta_{\mu \nu}T
    \label{1.77}
\end{align}
where $T_{\mu \nu}$ stands for the energy-momentum tensor of the matter fluids. Finally, by multiplying Eq. \eqref{1.77} by $\eta^{\mu \nu}$, we get
\begin{align}
    0 = - \frac{5}{4}\partial_{\lambda} \partial^{\lambda}h - 4m^2h + 2\partial^{\mu}\partial^{\nu}h_{\mu \nu}.
    \label{1.78}
\end{align}

In order to investigate Newton's law, we consider a point source at the origin and we look for a static solution of \eqref{1.78}. In the case of unimodular $F(R)$ gravity, there exist only three gauge degrees of freedom, due to the unimodular constraint \eqref{1.1}. By imposing, correspondingly, three gauge conditions, $\partial^{i}h_{ij} = 0$, Eq. \eqref{1.78} reduces to
\begin{align}
    -\frac{5}{4}\partial_k\partial^kh - 4m^2h = 0,
    \label{1.79}
\end{align}
and under a proper boundary condition, we obtain $h = 0$. By using the three gauge conditions and the equation $h = 0$, we can rewrite the components $(0,0), (i,j)$ and $(0,i)$ as 
\begin{align}
    \partial_i\partial^i h_{00} = \frac{3}{4}M\delta(\textbf{r}), \qquad \partial_k \partial^k h_{ij} = \frac{1}{4}M\delta(\textbf{r}), \qquad \partial_j\partial^jh_{0i}-\partial_i\partial^kh_{k0} = 0.
    \label{1.80}
\end{align}

Defining the Newtonian potential $\Phi$ by $h_{00}=2\Phi$, Eq. \eqref{1.80} gives the Poisson equation for the Newtonian potential $\Phi$, $\partial_i\partial^i = \frac{3k^2}{8}M\delta(\textbf{r})$. Hence, by redefining the gravitational constant $k$ as $\frac{3k^2}{4} \longrightarrow k^2 = 8 \pi G$, we do obtain the standard Poisson equation for the Newtonian potential, which solution is given by
\begin{align}
    \Phi = - \frac{G M}{r}.
    \label{1.81}
\end{align}
Thus, no correction to Newton's gravity law arises.

\subsection{Effective EOS for unimodular $F(R)$-gravity}

Let us consider the FRW metric, so the equations of motion \eqref{1.11} and \eqref{1.12} can be rewritten as,
\begin{align}
        0 = -\frac{1}{2}a^{-6}\left( F(R)-\lambda \right) + \left( 3\Dot{K} + 12K^2 \right)F_R - 3K \frac{d F_R}{d \tau} +\frac{1}{2}a^{-6}\rho_{matter},
    \label{1.82}
\end{align}
\begin{align}
        0= \frac{1}{2}a^{-6}\left( F(R)-\lambda \right) - \left( 6 K^2 + \Dot{K}\right)F_R + 5K \frac{d F_R}{d \tau} + \frac{d^2 F_R}{d \tau^2} + \frac{1}{2}a^{-6}p_{matter}.
    \label{1.83}
\end{align}
For convenience, we write $F(R)$ as the sum of the scalar curvature $R$ and the part which expresses the difference from the Einstein gravity case,
\begin{align}
    F(R) = R + f(R).
    \label{1.84}
\end{align}

We can now express the effective energy density $\rho_{eff}$ and also the effective pressure $p_{eff}$, including the contribution from $f(R)$ gravity, as follows
\begin{align}
    &\rho_{eff}=-\frac{1}{2}a^{-6}\left( f(R)-\lambda \right) + \left( 3\Dot{K} + 12K^2 \right)f_R - 3K \frac{d f_R}{d \tau} +\frac{1}{2}a^{-6}\rho_{matter} \\
    &p_{eff} = \frac{1}{2}a^{-6}\left( f(R)-\lambda \right) - \left( 6 K^2 + \Dot{K}\right)f_R + 5K \frac{d f_R}{d \tau} + \frac{d^2 f_R}{d \tau^2} + \frac{1}{2}a^{-6}p_{matter}
    \label{1.85}
\end{align}
which enables us to rewrite the equations \eqref{1.82} and \eqref{1.83} as in the Einsteinian case, namely
\begin{align}
    \rho_{eff} = 3K^2, \qquad p_{eff} = -\left(2\Dot{K} + 9K^2 \right).
    \label{1.86}
\end{align}
Now, we can find $\omega_{eff} = p_{eff}/\rho_{eff}$ and compare it to observational bounds. 

Let us consider the metric \eqref{1.24} with $H_1 = H_2 = H_0$, and the $F(R)$ given by \eqref{1.38} when $F_2 = 0$. By imposing $F_1 = \frac{7}{9}$, we can rewrite the function \eqref{1.38} and its derivative with respect to R, as
\begin{align}
    &F(R) = R + F_0 - \frac{R^2}{108H_0^2} = R + f(R), \nonumber \\
    &F_R(R) = 1 - \frac{R}{54H_0^2} = 1 + f_R(R),
    \label{1.87}
\end{align}
with the following Lagrange multiplier
\begin{align}
    \lambda(\tau) = \frac{138H_0^2}{9} + \frac{2C^2}{7H_0^2 \tau^2}-\frac{3C^4}{9604H_0^6\tau^4}.
    \label{1.88}
\end{align}
Using Eq. \eqref{1.36}, we can express $f(R)$ and $f_R(R)$ as functions of $\tau$, and finally write down $p_{eff}$ and $\rho_{eff}$, assuming $p_{matter} = \rho_{matter} = 0$, as
\begin{align}
    &p_{eff} = -\frac{C^2}{49 H_0^4 \tau^4}+\frac{\frac{9 C^2}{49 H_0^2 \tau^2}+12 H_0^2}{162 H_0^2 \tau^2}+\frac{\frac{3 C^4}{9604 H_0^6 \tau^4}-\frac{2 C^2}{7 H_0^2 \tau^2}-\frac{\left(\frac{9 C^2}{49 H_0^2 \tau^2}+12 H_0^2\right)^2}{108 H_0^2}+F_0-\frac{46 H_0^2}{3}}{2 H_0^2  \tau^2}   \nonumber \\
    &\rho_{eff} = -\frac{\frac{9 C^2}{49 H_0^2 \tau^2}+12 H_0^2}{162 H_0^2 \tau^2}-\frac{\frac{3 C^4}{9604 H_0^6 \tau^4}-\frac{2 C^2}{7 H_0^2 \tau^2}-\frac{\left(\frac{9 C^2}{49 H_0^2 \tau^2}+12 H_0^2\right)^2}{108 H_0^2}+F_0-\frac{46 H_0^2}{3}}{2 H_0^2 \tau^2}
\end{align}

And, finally, by taking into account the limit of $\omega_{eff}=\frac{p_{eff}}{\rho_{eff}}$ as $\tau \longrightarrow \infty$, it can be shown that
\begin{align}
    \lim_{\tau \longrightarrow \infty}\omega_{eff} = -1.
\end{align}
Hence, we have proven that such an anisotropic universe, in the course of isotropization, tends to realize the DE era.

\section{Conclusions and Future Perspectives}

This paper has been devoted to the study of the extension of unimodular Einsteinian gravity in the context of $F(R)$ gravity. Being more specific, we have applied the formalism of unimodular $F(R)$ gravity to describe a number of anisotropic evolution scenarios. In unimodular general relativity the determinant of the metric is constrained to be a fixed number or a function. It turns out, however, that the metric of a generic anisotropic universe is not compatible with this unimodular constraint, and thus we have been compelled to redefine the metric while taking into account the constraint properly. We have imposed the unimodular constraint to $F(R)$ gravity in the Jordan frame, by means of a Lagrangian multiplier, and derived the corresponding equations of motion. The resulting equations, being the result of the reconstruction method, allow us to determine which ones of the possible functions of the Ricci scalar are able to realize the desired evolution. 

By choosing several characteristic examples, we have shown explicitly how the reconstruction method works. The cosmological evolutions obtained in this way may prove to be important both for the pre-inflationary scenario and also for the late-time evolution of the universe. Indeed, we have considered, in particular, the specific de Sitter space-time realized in the context of unimodular $F(R)$ gravity, which can be suitable to describe both the early- and the late-time epochs of the universe history.

We have also stressed the potential value of using unimodular $F(R)$-gravity for defining a new effective equation of state, and explicitly compared the corrections it induces to Newton's law with  the corresponding ones in standard $F(R)$-gravity. 
We have shown that, while the corrections to Newton's law are in the standard $F(R)$-gravity case too big in some situations, and the model does not reproduce the correct dynamics in the Solar System, it turns out that, on the contrary, one can avoid this situation in the case of unimodular $F(R)$ gravity. For models in the framework of the unimodular $F(R)$ theory, we are actually able to reproduce the observed dynamics of the Solar System, in the limit of low curvature. 

A quite natural extension of this work will consist in the investigation of the specific effect of the anisotropies on the inflationary paradigm, in particular, under the form of noticeable changes in the Hubble slow-roll parameters. This will be reported elsewhere.

\section*{Acknowledgements}
This work has been partially supported by MICINN (Spain), project PID2019-104397GB-I00, of the Spanish State Research Agency program AEI/10.13039/501100011033, by the Catalan Government, AGAUR project 2017-SGR-247, and by the program Unidad de Excelencia María de Maeztu CEX2020-001058-M.

\end{document}